\documentclass[twocolumn]{aastex631}
\usepackage{bbding}
\usepackage{bm}

\newcommand{\msun}{{M}_{\sun}}
\usepackage{color}
\usepackage{lineno}
\usepackage{ulem}

\begin{document}

\title{First Observational Evidence for {an Interconnected Evolution between Time Lag and QPO Frequency among AGNs
}}
\correspondingauthor{Hao Liu, Yongquan Xue}
\email{liuhao1993@ustc.edu.cn, xuey@ustc.edu.cn}

\author[0009-0005-3916-1455]{Ruisong Xia}
\author[0000-0001-5525-0400]{Hao Liu\textsuperscript{\Envelope}}
\author[0000-0002-1935-8104]{Yongquan Xue\textsuperscript{\Envelope}}

\affiliation{CAS Key Laboratory for Research in Galaxies and Cosmology, Department of Astronomy, University of Science and Technology of China, Hefei 230026, China}
\affiliation{School of Astronomy and Space Science, University of Science and Technology of China, Hefei 230026, China}

\begin{abstract}

Quasi-periodic oscillations (QPOs) have been widely observed in black hole X-ray binaries (BHBs), which often exhibit significant X-ray variations.
Extensive research has explored the long-term evolution of the properties of QPOs in BHBs.
In contrast, such evolution in active galactic nuclei (AGNs) has remained largely unexplored due to limited observational data.
By using the 10 new \textit{XMM-Newton} observations for the narrow-line Seyfert 1 galaxy \hbox{RE J1034+396} from publicly available data, we analyze the characteristics of its X-ray QPOs and examine their long-term evolution.
The hard-band (1--4 keV) QPOs are found in all 10 observations and the frequency of these QPOs evolves ranging at $(2.47\text{--}2.83)\times10^{-4}\rm\ Hz$.
Furthermore, QPO signals in the soft (0.3--1 keV) and hard bands exhibit strong coherence, although, at times, the variations in the soft band lead those in the hard band (the hard-lag mode), while at other times, it is the reverse (the soft-lag mode).
The observations presented here serendipitously captured two ongoing {{lag reversals}} within about two weeks, which are first seen in \hbox{RE J1034+396} and also among all AGNs.
A transition in QPO frequency also takes place within a two-week timeframe, two weeks prior to its corresponding lag {{reversal}}, indicating a possible coherence between the transitions of QPO frequency and lag mode with {{delay}}.
The diagram of time lag versus QPO frequency clearly evidences this interconnected evolution {{with hysteresis}}, which is, for the first time, observed among AGNs.

\end{abstract}


\section{INTRODUCTION} \label{sec:intro}

The accretion processes involving black holes (BHs) play a pivotal role in powering both black hole X-ray binaries (BHBs) and active galactic nuclei (AGNs). These systems emit intense radiation spanning a broad wavelength range, from radio to X-ray bands. Since this emission originates from the inner regions of the accretion disk, X-ray observations have become a crucial window to investigate the physics of BH accretion. Variability and spectral characteristics in X-ray observations have been extensively analyzed in both BHBs and AGNs, among which the X-ray quasi-periodic oscillation (QPO) is a particularly intriguing phenomenon observed in both systems.

The discovery of QPOs began with the BHB \hbox{GX 339$-$4}, which exhibits strong X-ray variations with QPO frequencies ranging from tens of milli-{Hertz} to tens of {Hertz} \citep[][]{1979Natur.278..434S}. From then on, tens of BHBs were identified displaying X-ray QPOs, providing an ample dataset for studying QPO evolution during BHB outbursts. These QPOs were found to exhibit varying characteristics, suggesting an evolution tied to the accretion process \citep[][]{2005ApJ...629..403C}. Additionally, some studies have found that the phase lag between different energy bands evolves with the QPO frequency during outbursts in BHBs \citep[][]{2005ApJ...629..403C,2013ApJ...778..136P}. 

In contrast to BHBs that contain stellar mass BHs ($M_{\rm BH}\sim 10\msun$), AGNs, harboring supermassive black holes ($M_{\rm BH}\sim 10^6 \text{--} 10^{10}\msun$), were predicted to {also} exhibit QPOs. However, it was not until 2008 that the first AGN X-ray QPO was observed in the Seyfert galaxy \hbox{RE J1034+396} \citep[]{2008Natur.455..369G}. Owing to the massive central BH ($\sim 10^6 \text{--} 10^7\msun${;} \citealt{2016A&A...594A.102C}), the QPOs in \hbox{RE J1034+396} manifest themselves on much longer timescales than those of BHBs, around $\sim 3730 \pm 60$ seconds. Since the discovery of X-ray QPOs in this source, several other AGNs have also been found to exhibit QPOs, including 3C 273 \citep{2008ApJ...679..182E}, \hbox{1H 0707$-$495} \citep{2016ApJ...819L..19P}, \hbox{Mrk 766} \citep{2017ApJ...849....9Z}, \hbox{MS 2254.9$-$3712} \citep{2015MNRAS.449..467A}, {and \hbox{2XMM J123103.2+110648} \citep{2012ApJ...752..154T,2012ApJ...759L..16H,2013ApJ...776L..10L,2023MNRAS.523L..26K}}. Among these AGNs, \hbox{RE J1034+396} boasts the most significant QPO signals and the largest number of QPO observations, making it a prime candidate for in-depth study.

As a narrow-line Seyfert 1 galaxy located at {$z=0.043$}, \hbox{RE J1034+396} shows an extraordinarily steep spectrum in the soft X-ray band \citep[][]{1995MNRAS.276...20P} and its mass accretion rate approaches the Eddington limit \citep[][]{2016A&A...594A.102C}.
With a wealth of X-ray data available for RE J1034+396, numerous studies have been conducted to analyze its QPO properties, while the QPOs were only detected in six observations with \textit{XMM-Newton} spanning 11 years from May 2007 to October 2018 \citep[][]{2008Natur.455..369G,2014MNRAS.445L..16A,2020MNRAS.495.3538J}.
In the analysis of the power spectral densities (PSDs), the QPOs observed in \hbox{RE J1034+396} exhibit large quality factors and show some variation of the QPO frequency, reminiscent of the high-frequency QPOs observed at 35 or 67 Hz in the BHB \hbox{GRS 1915+105} \citep[see ][]{2013MNRAS.435.2132M, 2014MNRAS.445L..16A,2021MNRAS.500.2475J}.
The long-term evolution of QPO properties has been observed in BHBs. Similar behavior could potentially be found in AGNs as well.
The QPO period of \hbox{RE J1034+396} was observed to decrease by approximately 250 seconds in 11 years, as reported by \cite{2020MNRAS.495.3538J}. Furthermore, they noted a reversal in the phase lag between the hard (1--4 keV) {and} soft (0.3--1 keV) X-ray light {curves around} the QPO frequency, contrasting with previous findings. These results suggest that AGN QPOs may indeed evolve over time, similar to what has been observed in BHBs. However, the characteristics of this evolution still remain unclear.

In this letter, we present a study based on 10 new \textit{XMM-Newton} observations {{of \hbox{RE J1034+396}}} obtained from the publicly available data. These observations span a period of half a year, enabling us to analyze the evolution of QPO properties on timescales of weeks to months. 
This letter is organized as follows. We present the details of the observations and data reduction in Section~\ref{sec:obs}. The subsequent sections cover a comprehensive analysis of PSD and QPO properties, as well as an examination of time lags in the vicinity of the QPO frequency (Section~\ref{sec:qpo}). We discuss our findings in Section~\ref{sec:dis} and conclude in Section~\ref{sec:conc}.

\section{OBSERVATIONS AND DATA REDUCTION} \label{sec:obs}

\begin{figure*}
    \centering
    \includegraphics[scale=0.64]{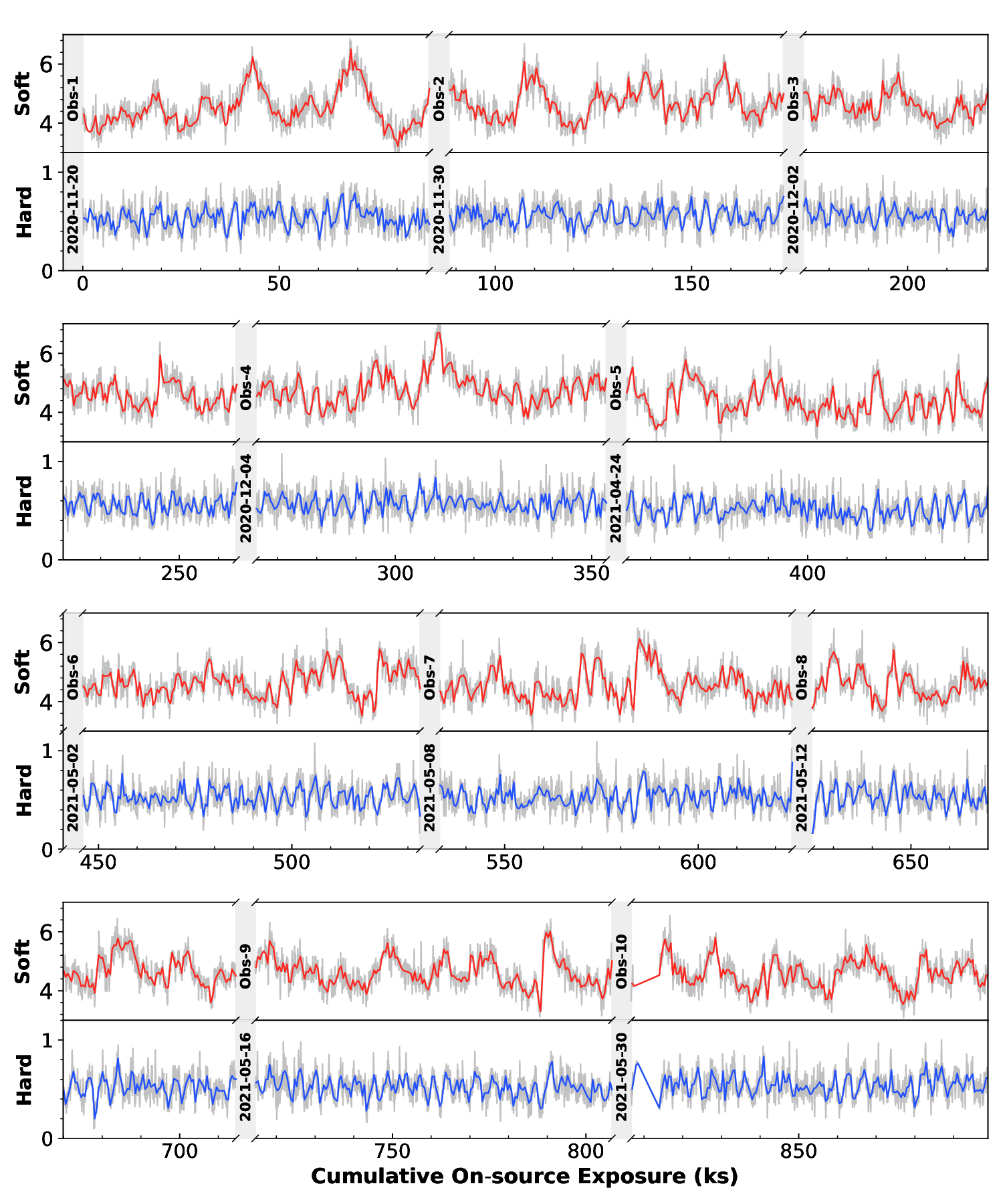}
    \caption{Light curves for the 10 observations in the soft (0.3--1 keV) and hard (1--4 keV) bands, separated by gray vertical intervals with the observation date and number annotated at the beginning of each light curve. The gray curves represent data binned to 100~s with typical uncertainty less than 0.01 counts/s (not shown), while the red and blue curves are binned to 500~s. The light curves, interpolated where being affected by background flares, have nearly identical average count rates.}
    \label{fig:lc}
\end{figure*}

RE J1034+396 has been observed using the \hbox{\textit{XMM-Newton}} \citep[][]{2001A&A...365L...1J} satellite, with a record of more than 20 observations to date. Notably, a substantial portion of the publicly available data stems from observations conducted between November 2020 and June 2021. This abundance of data offers a unique opportunity to study the X-ray variability of \hbox{RE J1034+396} over this temporal span. For this analysis, we select the 10 new observations as listed in Table~\ref{Tab:tab}. 
Each observation has an observing time around 90 kilo-seconds. Our study focuses on analyzing the timing properties of the source, and therefore, we exclusively collect data from the European Photon Imaging Cameras (EPIC). All the data are reduced using the Science Analysis Software (SAS, version 20.0.0) with the latest calibration files. The calibrated EPIC event files are generated from the original observation data files using the \textsc{epproc} task.
{We extract the event files} following the standard data reduction procedure, wherein only single and double pixel events are considered ($\rm PATTERN \le 4$ and $\rm FLAG=0$).
Moreover, a filter condition that $\rm RATE \le 0.4$ within the 10--12 keV light curve is applied to extract {{a}} good-time-interval (GTI) file. 
The source regions are mainly selected within a $40\arcsec$ radius circle encompassing the source of interest, while the background regions are selected as a nearby source-free circle with the same size. 

To ensure consistency in data handling, we exclusively utilize the EPIC-pn data for all the observations. The extraction of source and background light curves is executed using the \textsc{evselect} task, each with distinct region selections, and the final background-subtracted light curves are produced by the \textsc{epiclccorr} task. For the analysis of the light curves, we set the binning time to 100 seconds. To facilitate the analysis of time lags between different energy bands, we extract light curves in two broad bands, i.e., the soft band (0.3--1 keV) and the hard band (1--4 keV). The average count rate of each observation is listed in Column~{(7)} of Table~\ref{Tab:tab} and the typical uncertainty is less than $\rm 0.01\ counts\ s^{-1}$. We show the light curves of the recent 10 observations in Figure~\ref{fig:lc}. The stochastic variability in the soft band is more pronounced than in the hard band, while the QPO in the hard band is visually more prominent.

\begin{table*}
\begin{center}
\caption{Information of \textit{XMM-Newton} Observations and the Parameters of PSD Modeling Results of RE J1034+396. }\label{Tab:tab}
 \renewcommand{\arraystretch}{1.3}
\setlength{\tabcolsep}{1.5mm}{}
\begin{tabular}{cccccccccccc}
\hline
\hline
Obs. No. & ObsID & Obs. Date & GTI & {HR} & Band &Rate   & $R_{\rm QPO}$ & Sig. &{$ \rm RMS_{QPO}$}& $f_{\rm QPO}$ & $Q_{\rm QPO}$\\
&&(yyyy-mm-dd)&($\rm ks$)&&&($\rm counts\ s^{-1}$)&&($\sigma$)&{(\%)}&($\rm \times 10^{-4}\ Hz$)& \\
(1)&(2)&(3)&(4)&(5)&(6)&(7)&(8)&(9)&(10)&(11)&(12) \\
\hline
   Obs-1 & 0865010101 & 2020-11-20 & $86.4 $&{$0.108$}&soft& $4.48  $ & $6.6$ & $2.1 $ & --- &  --- &  ---\\
   & & & &&hard& $0.54  $ & $43.3$ & $ 6.3 $ &{$ 10.04$}& {$ 2.56\pm 0.05$} &  {$ 17.6$} \\
   Obs-2 & 0865011001 & 2020-11-30 & $85.2 $&{$0.106$}&soft& $4.70  $ & $12.6$ & $3.1 $ & --- & --- &  ---\\
   & & & &&hard& $0.56  $ & $51.8$ & $ 6.9 $ &{$ 10.98$}& {$ 2.47\pm 0.03$} & {$ 12.3$} \\
   Obs-3 & 0865011201 & 2020-12-02 & $89.2 $ &{$0.109$}&soft& $4.58  $ & $32.3$ & $ 5.3 $ &{$ 3.28$}& {$ 2.58\pm 0.05 $} & {$ 19.6$} \\
   & & & &&hard& $0.56  $ & {$ 81.4$} & {$  >9.0 $} &{$ 9.26$} & {$ 2.57\pm0.05$} & {$ 19.1$} \\
   Obs-4 & 0865011101 & 2020-12-04 & $91.0 $ &{$0.107$}&soft& $4.78  $ & $69.6$ & $ 8.1 $ &{$ 3.43$}& {$ 2.53\pm0.03$}& {$ 21.5$} \\
   & & & &&hard& $0.57  $ & $106.4$ & $ >9.0 $ &{$ 10.48$}& {$ 2.57\pm0.02$} & {$ 13.5$} \\
   Obs-5 & 0865011301 & 2021-04-24 & $91.4 $ &{$0.105$}&soft& $4.37  $ & $52.8$ & $ 7.0 $ &{$ 3.48$}& {$ 2.88\pm0.03$}& {$ 25.6$} \\
   & & & &&hard& $0.51  $ & $79.6$ & $ >9.0 $ &{$ 13.81$}& {$ 2.83\pm0.05$} & {$ 15.9$} \\
   Obs-6 & 0865011401 & 2021-05-02 & $87.2 $ &{$0.103$}&soft& $4.63  $ & $24.9$ & $ 4.6 $ &{$ 3.59$}& {$ 2.71\pm0.06$} & {$ 19.2$} \\
   & & & &&hard& $0.53  $ & $103.4$ & $ >9.0 $ &{$ 13.15$}& {$ 2.74\pm0.05$} & {$ 17.6$} \\
   Obs-7 & 0865011501 & 2021-05-08 & $91.4 $ &{$0.105$}&soft& $4.54  $ & $26.2$ & $ 4.7 $ &{$ 3.35$}& {$ 2.62\pm0.05$} &  {$ 18.6$} \\
   & & & &&hard& $0.53  $ & $65.0$ & $ 7.8 $ &{$ 12.03$}& {$ 2.60\pm0.05$}&  {$ 17.8$} \\
   Obs-8 & 0865011601 & 2021-05-12 & $88.8 $ &{$0.101$}&soft& $4.63  $ & $30.7$ & $ 5.2 $ &{$ 1.95$}& {$ 2.62\pm0.03$}&  {$ 23.2$} \\
   & & & &&hard& $0.52  $ & $133.2$ & $ >9.0 $&{$ 11.04$} & {$ 2.61\pm0.03$} &  {$ 17.4$} \\
   Obs-9 & 0865011701 & 2021-05-16 & $91.5 $ &{$0.101$}&soft& $4.61  $ & $13.7$ & $3.3 $ & --- & --- &  ---\\
   & & & &&hard& $0.52  $ & $71.5$ & $ 8.2 $ &{$ 15.16$}& {$ 2.63\pm0.06$} &  {$ 16.8$} \\
   Obs-10 & 0865011801 & 2021-05-30 & $85.5$ &{$ 0.106$}&soft& $4.57  $ & $14.0$ & $3.3 $ & ---  & --- &  ---\\
   & & & &&hard& $0.54  $ & $95.3$ & $ >9.0 $&{$ 13.38$} & {$ 2.52\pm0.06$} &  {$ 16.0$} \\
\hline
\hline
\end{tabular}

\tablecomments{Columns are as follows: 
(1) The observation number in this letter; 
(2) The observation ID; 
(3) The observation date; 
(4) The good time interval in EPIC-pn; 
(5) {The hardness ratio $\rm HR=H/(S+H)$, H for the hard-band rate and S for the soft;} 
(6) {The soft band (0.3--1 keV) or the hard band (1--4 keV)}; 
(7) The count rate, {exhibiting typical uncertainties} less than $\rm 0.01\ counts\ s^{-1}$; 
(8) The 2$\times$data/continuum value at the QPO frequency{, used to assess the significance by comparing it with a $\chi^2$ distribution}; 
(9) The significance of the observed QPO; 
{(10) The fractional RMS of the QPO obtained through integration of the best-fit} \textsc{lorentzian} profile, with typical uncertainties of $1.5\%\ (4.7\%)$ for the soft (hard) band; 
(11) The fitted QPO frequency of the best-fit QPO \textsc{lorentzian} profile; 
(12) The quality factor for the QPO signal. The dashes in the table imply that the QPO signals lack significance for analysis.}
\end{center}
\end{table*}

\section{ANALYSES AND RESULTS} \label{sec:qpo}

\subsection{X-ray PSDs}

\begin{figure*}
    \centering
    \includegraphics[scale=0.48]{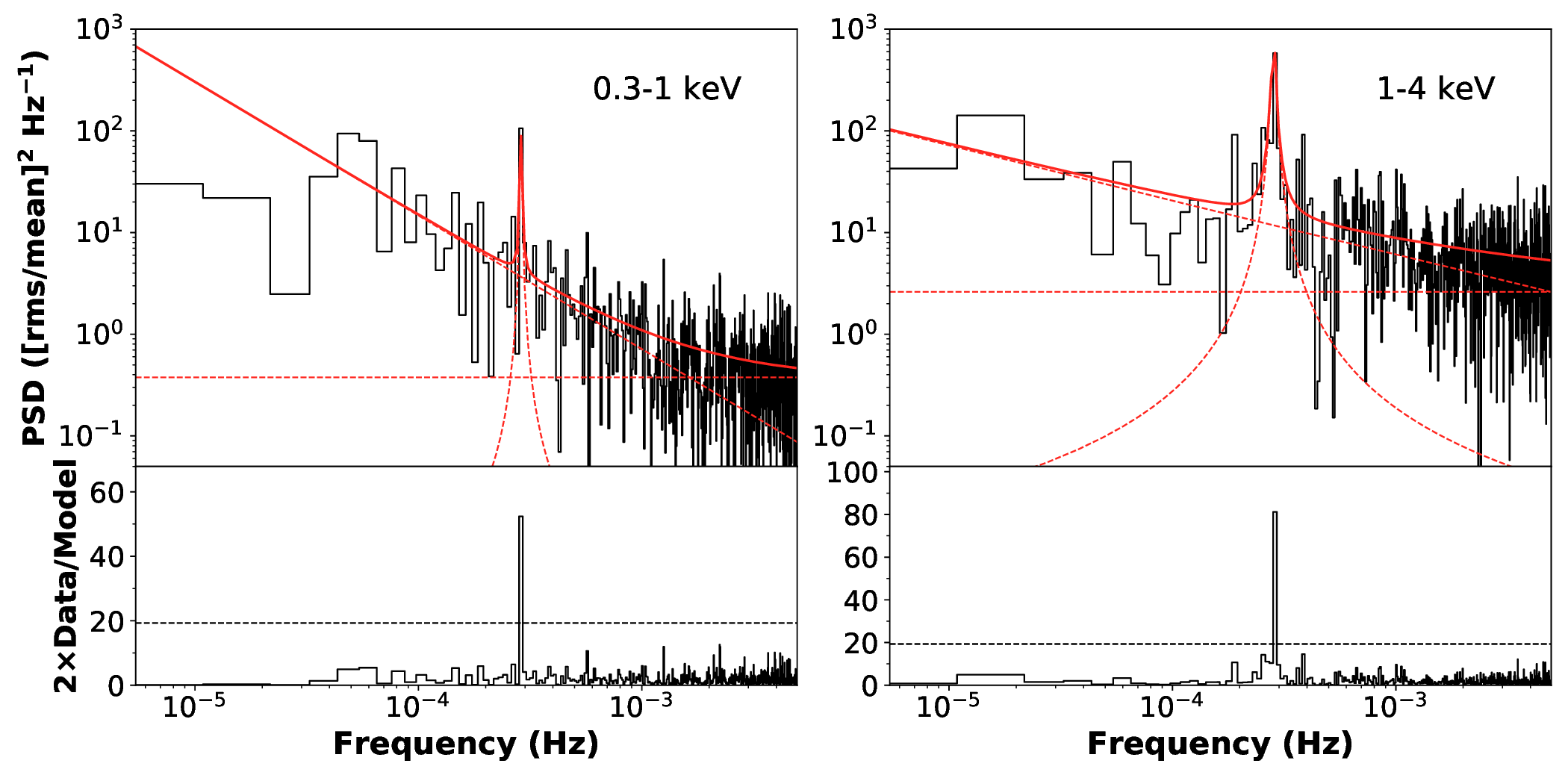}
    \caption{PSDs of Obs-5 for \hbox{RE J1034+396} in the soft band (0.3--1 keV) and the hard band (1--4 keV), respectively. In the upper panels, the red solid curves represent the best-fit model, and the red dashed curves depict the components of a power law, a constant Poisson noise, and a \textsc{Lorentzian} profile for the QPO. The bottom panels illustrate the ratios of the observed PSDs to the PSD continuum models; the black dashed lines represent the $4 \sigma$ confidence limits for red-noise fluctuations close to the QPO frequency.}
    \label{fig:PSD}
\end{figure*}

To investigate the significance and properties of the QPO signals, we transform the X-ray light curves into the frequency domain, i.e., the PSD diagram. We conduct our PSD analysis using the Lomb-Scargle periodogram technique \citep{1976Ap&SS..39..447L, 1982ApJ...263..835S} with RMS normalization \citep[][]{1990A&A...230..103B} such that the unit of PSD is $\rm (rms)^2/Hz$ and the noise is equal to $\rm 2/\langle rate\rangle $ {\citep{2003MNRAS.345.1271V}}. 
We illustrate our analysis of examining the PSDs of the {soft and hard} bands in Figure~\ref{fig:PSD}, utilizing data from \hbox{Obs-5} as an example.
In most cases, QPO signals are prominently visible in the PSDs, with the significance tested in Section~\ref{subsec:sig}, exhibiting similar characteristics (estimated in Section~\ref{subsec:pro}) to those previously reported in the literature, thus indicating consistent QPO activity as documented in \citet{2014MNRAS.445L..16A} and \citet{2020MNRAS.495.3538J}. Notably, we find evolution in the characteristics of the QPO signals, prompting a more in-depth investigation into the variability of these signals, which will be discussed in Section~\ref{sec:dis}.

\subsection{Significance of QPO Signals}\label{subsec:sig}

We assess the significance level of the QPO signals by testing a continuum-only hypothesis \citep[][]{2010MNRAS.402..307V,2020MNRAS.495.3538J}. We fit the PSD with a model of continuum consisting only of \textsc{powerlaw + constant} components using the maximum likelihood estimates (MLE) methods \citep[][]{2010MNRAS.402..307V}, of which the likelihood is 
\begin{equation}
    \log\ p({\bm I}|{\bm \theta},H) = - \sum\limits_{j=1}^N{\frac{I_j}{S_j}+\log\ S_j},
\end{equation}
known as the Whittle likelihood method \citep[][]{1953ArM.....2..423W}, where ${\bm I}$ corresponds to the observed PSD, ${\bm \theta}$ denotes the parameters of the model, $H$ signifies the hypothesis, ${I_j}$ represents the $j$-th data in $\bm I$, and $S_j$ represents the $j$-th value of the fitted continuum model, respectively. 

By defining $R_j=2I_j/S_j$, we identify the highest data/model outlier {${\rm max}(R_j)$} close to the QPO frequency{, denoted as $R_{\rm QPO}$ (Column~(8) in Table~\ref{Tab:tab}).} {For simplicity, the significance of the QPO signal (listed in Column~(9) of Table~\ref{Tab:tab}) is evaluated by comparing the value of {$R_{\rm QPO}$} to a standard $\chi^2$ distribution with 2 degrees of freedom {\citep{2008Natur.455..369G}}.} 
We select a $4\sigma$ threshold to determine the presence of a QPO signal, a choice substantiated by the results of fitting the QPO {signals} (Section~\ref{subsec:pro}), as those with significance $\rm \le 4\sigma$ are hardly fitted by the model.
The $R$ value for a $4\sigma$ significance level is 19.3 in a standard $\chi^2$ distribution with 2 degrees of freedom, which is close to the values obtained from simulation tests conducted by \cite{2020MNRAS.495.3538J}.

QPO signals are observed in the hard band across all the 10 observations, while 4 observations show no significant QPO signals in the soft band. 
The QPO signals in the hard band exhibit larger significance compared to the soft band, consistent with the findings based on earlier observations \citep{2014MNRAS.445L..16A}.

\subsection{Properties of QPO Signals}\label{subsec:pro}
\begin{figure*}[!ht]
    \centering
    \includegraphics[scale=0.66]{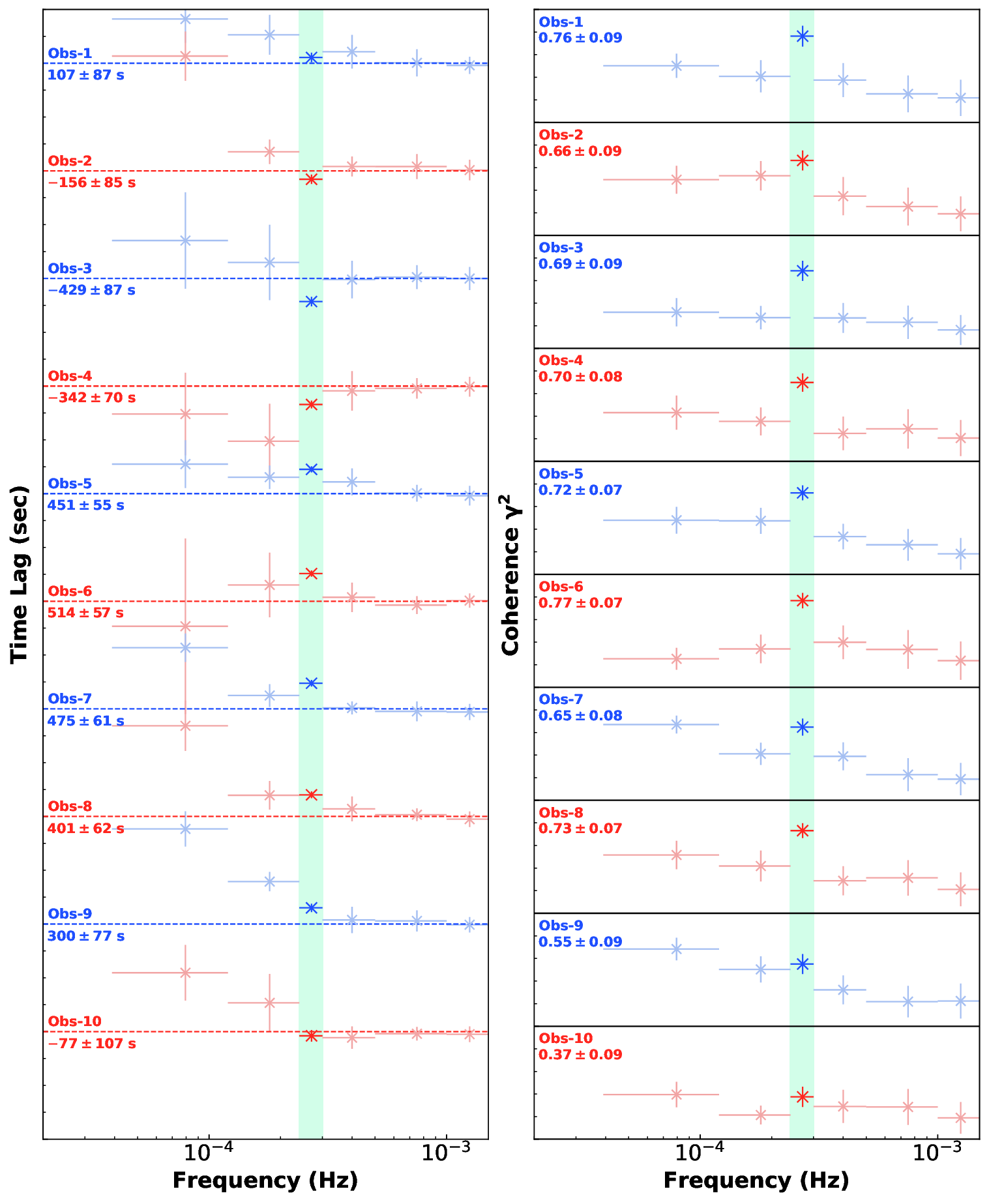}
    \caption{Time lag (left) and coherence (right) as a function of frequency between the soft (0.3--1 keV) and hard (1--4 keV) bands. 
    In the left panel, the horizontal dashed lines denote 0 values of time lag, and positive values relative to these dashed lines represent hard-band lags. Red and blue markers are interchangeably used to distinguish different observations. The spacing between the dashed lines in the left panel measures 2000 seconds, while the interval between the solid lines in the right panel corresponds to a coherence of unity. The frequency range of $(2.{4}\text{--}3.0)\times 10^{-4}\ \rm Hz$ close to the QPO frequency is highlighted in green and the values estimated in this frequency range are noted on the left side, indicating the QPO time lag and coherence. The error bars for both the time lag and the coherence are estimated by bootstrapping.}
    \label{fig:lag-f}
\end{figure*}

\begin{figure*}[!ht]
    \centering
    \includegraphics[scale=0.75]{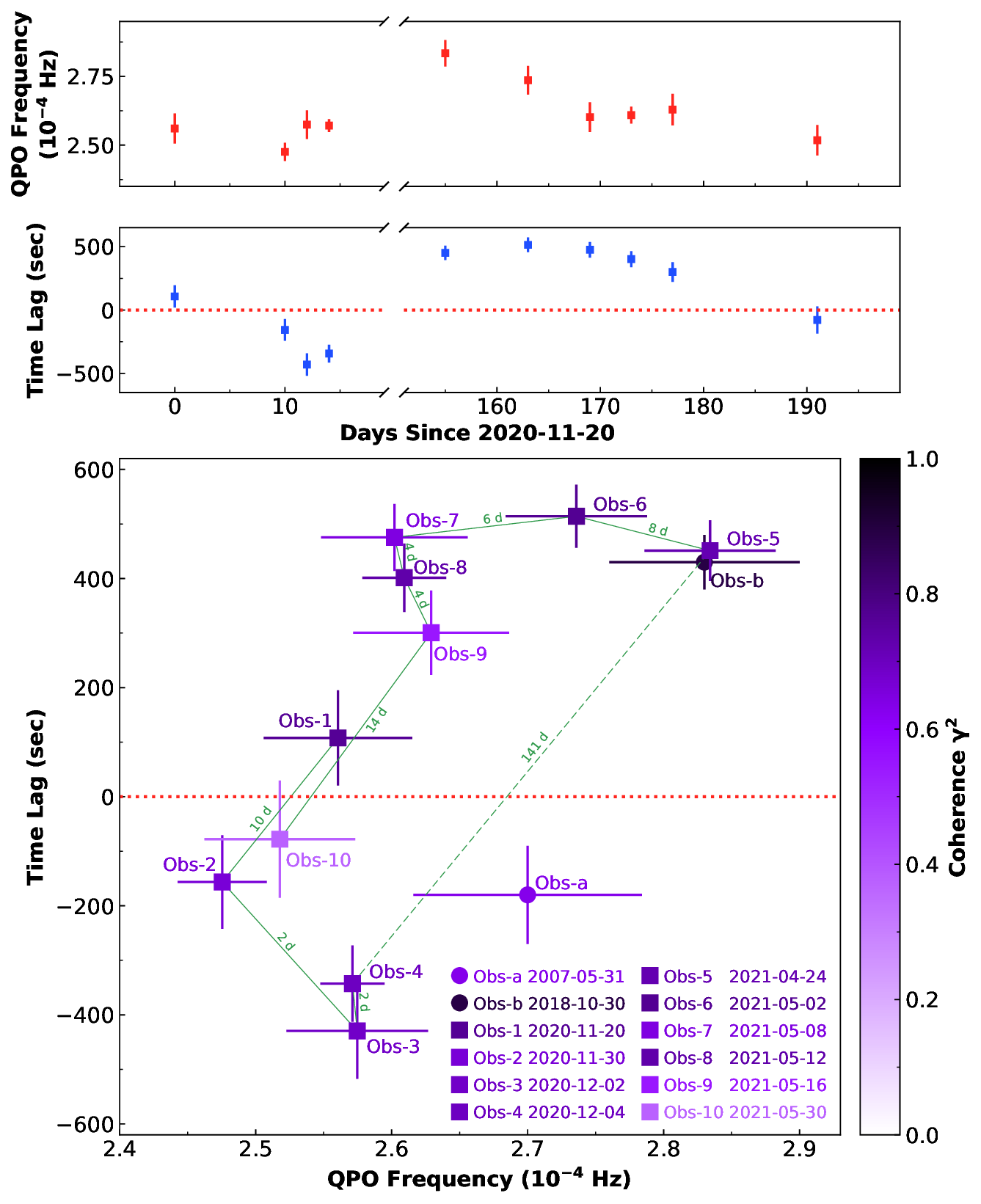}
    \caption{The evolution of hard-band $f_{\rm QPO}$ and QPO time lag (measured around $f_{\rm QPO}$) between the soft (0.3--1 keV) and hard (1--4 keV) bands. Top panel: $f_{\rm QPO}$ and the time lag as a function of time. Bottom panel: $f_{\rm QPO}$ vs. time lag. Ten observations analyzed in this work are denoted by squares. \hbox{Obs-a} and \hbox{Obs-b} (circles) are the previous observations showing QPO signals and the time lags, derived from \cite{2008Natur.455..369G} and \cite{2020MNRAS.495.3538J}. Colors are assigned based on the coherence between the variability around $f_{\rm QPO}$ of the soft and hard bands. The horizontal dotted line marks 0 values of time lag. The intervals between observations in units of days are annotated in green.}
    \label{fig:lag-evo}
\end{figure*}

Markov chain Monte Carlo (MCMC) approximations are widely employed for their efficiency in estimating model parameters. In our analysis, we utilize \textsc{emcee} to sample the joint posterior probability {densities} of all parameters in the model, encompassing the \hbox{\textsc{powerlaw + lorentzian + constant}} components, producing the {posteriors} of both the parameters of the model and the predicted model PSD. 
The upper panels of Figure~\ref{fig:PSD} display the PSDs for the soft and hard bands of \hbox{Obs-5}, along with the corresponding MCMC fitting results represented by red curves.
We quantify the fractional root mean square (RMS) of the QPO (listed in Column~(10) of Table~\ref{Tab:tab}) by integrating the best-fit \textsc{lorentzian} profile \citep{2020MNRAS.495.3538J}.
The estimate of the QPO frequency ($f_{\rm QPO}$) for each observation is shown in Column~(11) of Table~\ref{Tab:tab}, which is consistent with the frequency where the \textsc{lorentzian} component peaks.
Additionally, we compute the full width at half maximum (FWHM) of the \textsc{lorentzian} profile ($W_{\rm QPO}$) and calculate the quality factor of the signal $Q_{\rm QPO}=f_{\rm QPO}/W_{\rm QPO}$, as shown in Column~(12) of Table~\ref{Tab:tab}.

We find that $f_{\rm QPO}$ is generally decreasing across the six observations in 2021 (\hbox{Obs-5} to \hbox{Obs-10}), which peaks at \hbox{Obs-5}, reaching $2.88\times 10^{-4}\ \rm Hz$ in the soft band and $2.83\times 10^{-4}\ \rm Hz$ in the hard band, respectively. The evolution of hard-band $f_{\rm QPO}$ as a function of time is shown in the top panel of Figure~\ref{fig:lag-evo}.

\subsection{Time Lag and Coherence} \label{sec:lag}

To derive the time lag between light curves in the soft and hard bands of a specific observation, we compute the cross-power spectral density (CSD) with the \textsc{scipy} package, which is equal to Equation (9) in \cite{2014A&ARv..22...72U}. Subsequently, we conduct a phase lag analysis based on this CSD, which corresponds to the time lag in the time domain \citep[][]{2020MNRAS.495.3538J}. The left panel of Figure~\ref{fig:lag-f} shows the time lag as a function of frequency, and the region close to the QPO frequency ($\rm (2.{4}\text{--}3.0) \times 10^{-4}\ Hz$), in which the QPO lag values are estimated, is highlighted.
The QPO lags of \hbox{Obs-1}, \hbox{Obs-5}, \hbox{Obs-6}, \hbox{Obs-7}, \hbox{Obs-8}, and \hbox{Obs-9} predominantly exhibit positive values, also shown as a time series in Figure~\ref{fig:lag-evo}, suggesting a hard-lag mode where variations in the soft band precede those in the hard band. Conversely, the lag mode shifts to the reverse in \hbox{Obs-2}, \hbox{Obs-3}, \hbox{Obs-4}, and \hbox{Obs-10}, indicating that the hard band leads the variations (i.e., the soft-lag mode).
A conspicuous lag mode reversal occurs between \hbox{Obs-1} and \hbox{Obs-2}, merely {10 days} apart in the timeline, and a similar reversal is also observed between \hbox{Obs-9} and \hbox{Obs-10} with two weeks apart and albeit with lower coherence (also see Figure~\ref{fig:lag-evo}).

The coherence $\gamma^2$ between the hard and soft bands \citep[][]{2014A&ARv..22...72U} serves as a valuable indicator for testing whether the QPO signals in different energy bands correspond to each other. 
As shown in the right panel of Figure~\ref{fig:lag-f}, the coherence values are mostly $\sim$0.7 around the QPO frequency, indicative of a strong relationship between signals within this frequency range.
Typically, the coherence near the QPO frequency tends to be highest compared to frequencies across a broader range, with two exceptions in \hbox{Obs-9} ($\gamma^2 = 0.55$) and \hbox{Obs-10} ($\gamma^2 = 0.38$), where the coherence appears to decrease in this period, suggesting that the stochastic variability gradually dominates the soft-band light curve. 
The top panel of Figure~\ref{fig:lag-evo} displays the temporal evolution of the QPO time lag.

\section{DISCUSSIONS} \label{sec:dis}

\subsection{Interconnected Evolution of QPO Frequency and Time Lag}
The long-term variation of $f_{\rm QPO}$ of \hbox{RE J1034+396} has been reported by \citet{2020MNRAS.495.3538J} based on earlier observations, which varies from $(2.5\text{--}2.7)\times10^{-4}\rm\ Hz$ around 2010 \citep{2014MNRAS.445L..16A} to $2.83\times10^{-4}\rm\ Hz$ in 2018 \citep{2020MNRAS.495.3538J}. Our analysis of the 10 new observations suggests that the variation of $f_{\rm QPO}$ in the hard band of \hbox{RE J1034+396} may not occur over the previously assumed timescales of years. Instead, it is found that $f_{\rm QPO}$ decreases from $2.83\times10^{-4}\rm\ Hz$ (Obs-5) to $2.60\times10^{-4}\rm\ Hz$ (Obs-7) within a span of two weeks. However, how long it stays in either a low- or high-frequency state is still unknown.
Previous works have reported {{a reversal between the}} two time-lag modes. 
In the {2007 observation (\hbox{ObsID~0506440101}, hereafter \hbox{Obs-a}; \citealt{2008Natur.455..369G,2011MNRAS.417..250M,2011MNRAS.418.2642Z})}, the QPO in the hard band was found to lead the variations. 
Conversely, in the 2018 observation \citep[\hbox{ObsID~0824030101}, hereafter \hbox{Obs-b};][]{2020MNRAS.495.3538J}, the QPO in the soft band led the variations with a high coherence \citep{2021MNRAS.500.2475J}. It is worth noting that in Obs-b under the hard-lag mode, $f_{\rm QPO}$ was also higher compared to the other earlier observations.

In this study, our analysis of the 10 observations reveals the presence of both the soft-lag and hard-lag modes within a span of half a year.
We present the relation between the evolution of $f_{\rm QPO}$ and time lag in the bottom panel of Figure~\ref{fig:lag-evo}.
A similar trend to previous works is observed, albeit with richer information. Specifically, observations with higher $f_{\rm QPO}$ values (i.e., \hbox{Obs-5}, \hbox{Obs-6}, and \hbox{Obs-b}) exhibit a hard lag.
The QPO frequency transition takes place between \hbox{Obs-5} and \hbox{Obs-7}, while the lag {{reversal}} occurs from \hbox{Obs-7} to \hbox{Obs-10}, with a two-week {{delay}}. 
The lag reversal recurs (i.e., Obs-9 to Obs-10, Obs-4 to Obs-5, and Obs-1 to Obs-2, as well as Obs-a to Obs-b in the previous observations), and the evolution of $f_{\rm QPO}$ and time lag is closely interconnected with {{hysteresis}}. We therefore speculate a counter-clockwise cyclic evolution of $f_{\rm QPO}$ and time lag, seemingly demonstrated in Figure~\ref{fig:lag-evo}. 

Unfortunately, the limited number of observations for this source makes it challenging to firmly establish this potential cycle. It is also difficult to determine precise time scales for the transitions or ascertain if there is a cyclical pattern of evolution. To validate these conjectures, more intensive and higher-cadence X-ray observations for this source would be necessary.

\subsection{Comparison with BHB Systems}

Before \citet{2020MNRAS.495.3538J} and this work, the QPO lag reversal has solely been seen in BHBs.
Early in 2000, \cite{2000ApJ...541..883R} firstly found the X-ray QPO phase lag reversal from the hard-lag to soft-lag mode in the BHB GRS 1915+105, associated with the low-frequency QPO increasing from 0.5 to 10~Hz, which is in contrast to the trend seen in Figure~\ref{fig:lag-evo} for RE J1034+396. 
\citet{2017MNRAS.466..564G} characterized a group of BHBs including \hbox{H 1743$–$322}, \hbox{XTE J1859+226}, \hbox{XTE J1550$–$564}, \hbox{XTE J1817$–$330}, and \hbox{MAXI J1659$–$152} in their global study of type-B QPOs.
The QPOs observed in these BHBs exhibit a decreasing hard time lag, followed by a reversal towards a soft time lag. This transition accompanies a softening of the energy spectrum, a trend contrasting the results in this study, where the spectrum exhibits the highest hardness (see Column~(5) in Table~\ref{Tab:tab}) during \hbox{Obs-3} (in the soft-lag mode).
A QPO lag reversal was also observed during the transition from type-C to type-B for the BHB \hbox{MAXI J1820+070} \citep{2023MNRAS.525..854M}, 
with the variation of these QPO frequencies (approximately $1\ {\rm dex}$) being considerably larger than that in RE J1034+396.

It has been noted that the high-frequency QPO at 67 Hz (rather than the aforementioned low-frequency QPO) in \hbox{GRS 1915+105} is consistent with the QPO found in RE J1034+396 under the BH mass scaling, given their shared similarities. Both QPOs exhibit large quality factors and small but significant QPO frequency variations around $5\%$ \citep[see][]{2009MNRAS.394..250M, 2014MNRAS.445L..16A, 2019MNRAS.489.1037B, 2021MNRAS.500.2475J}. 
\cite{2021MNRAS.500.2475J} found the hard-lag mode of \hbox{RE J1034+396} observed in 2018, similar to the 67-Hz QPO of \hbox{GRS 1915+105} whose
reverse mode (i.e., the soft-lag mode) was not observed until its recent detection \citep{2024MNRAS.527.4739M}, providing new evidence for the similarity.

In searching for an explanation for the QPO evolution of \hbox{RE J1034+396}, we view ideas proposed for the lag variation in BHBs.
The first definite model for the accretion flow in \hbox{GRS 1915+105} was presented by \citet{2000ApJ...538L.137N}. 
They sketched a corona consisting of hot and warm regions. The QPO originates at the inner disk edge, and the lag reversal stems from the varying size of the corona.
A comparable Comptonization model was proposed and successfully explained the QPO frequency and lag evolution for low-frequency QPOs in \hbox{GRS 1915+105} \citep{2021MNRAS.503.5522K}.
Similarly, a stratified model assuming propagating fluctuations from the disk to the corona \citep{2023arXiv231208302U} may also help with understanding the origin of the QPO lag evolution.

Spectral models in AGNs support a stratified structure \citep{2018MNRAS.480.1247K}.
By fitting the X-ray spectra of \hbox{RE J1034+396}, two or three thermal components are considered to contribute to the emission \citep[][]{2010ApJ...718..551M,2021MNRAS.500.2475J}. 
However, the interconnected evolution of the QPO frequency and time lag with {{hysteresis}} is hard to interpret by a geometry-varying origin, demanding quasi-simultaneous transitions.
One possible explanation to mitigate this issue is to assume some orbiting intrinsic oscillator moving between the stratified components in the accretion flow. 
The variation of $f_{\rm QPO}$ can be attributed to the Doppler effect, while the observed two-week {{delay}} of the QPO time lag reversal potentially arises from the movement of this oscillator in an elliptical orbit observed along some specific viewing angle.
The origin of such an oscillator is worth further exploration in AGN systems. The accretion from a white dwarf in a very eccentric orbit around the central massive black hole \citep{2023MNRAS.523L..26K} could be one of the possibilities.

\subsection{{Other Remarks}}

It is worth noting that in \hbox{Obs-1}, \hbox{Obs-2}, \hbox{Obs-9}, and \hbox{Obs-10}, the QPOs in the soft band are too weak to detect. And the soft-band $\rm RMS_{QPO}$ is generally higher when larger absolute lag values are measured, possibly indicating the weakening of soft-band QPOs during a transitional state of {{lag reversal}}.
The presence of QPO in RE J1034+396 is thought to be associated with spectral characteristics \citep[][]{2014MNRAS.445L..16A}, where the observations harboring QPOs generally have lower soft X-ray fluxes and higher X-ray hardness. 
In contrast, the observations analyzed in this work show minimal variations in count rates and consistently maintain high hardness ({see} Table~\ref{Tab:tab}). The QPO signals are observed in the hard band of all 10 observations, with the spectra being almost the same. Therefore, the detection of the QPO might not be linked to either the intensity or hardness.
A detailed analysis based on the spectra will be conducted in follow-up works to confirm these findings.

\section{CONCLUSIONS} \label{sec:conc}

With the recent addition of 10 publicly available \textit{XMM-Newton} observations of RE J1034+396, we now have the opportunity to further explore the long-term evolution of its QPOs. Previously, only 6 \textit{XMM-Newton} observations were identified with QPOs. These earlier observations suggested that the QPO exhibited an intrinsic hard lag, while the soft-lag mode was attributed to a stochastic process \citep{2020MNRAS.495.3538J}. However, in this study, we demonstrate the existence of the two distinct lag modes with high coherence and analyze the temporal evolution of the QPO properties. The primary findings and conclusions are as follows:
\begin{enumerate}
    \item QPO signals are consistently present in the hard band in all the 10 new observations of RE J1034+396, with the QPO frequency evolving within the range of $(2.47 \text{--}2.83)\times10^{-4}\rm\ Hz$.
    \item The QPO time lag between the soft and hard bands exhibits two modes: the hard-lag mode in which the soft band leads the QPO and the soft-lag mode in which the hard band leads. Higher QPO frequency seems to be associated with the hard-lag mode. 
    \item We report two ongoing {{lag reversals}} spanning a period of about two weeks.
    A transition in QPO frequency also takes place within a two-week timeframe, two weeks prior to its corresponding lag {{reversal}}, indicating a possible coherence between the transitions of QPO frequency and lag mode with {{delay}}. The diagram of time lag versus QPO frequency clearly evidences this interconnected evolution {{with hysteresis}}, which is, for the first time, observed among AGNs.
    \item The recurrence of the two lag modes and the interconnected evolution suggest a likely counter-clockwise cyclic evolution in the time lag-QPO frequency diagram, potentially exhibiting periodic characteristics over an extended period of time.
    
\end{enumerate}

\begin{acknowledgments}
We thank the anonymous referee for providing valuable comments that significantly improved this work. This work is based on observations obtained with XMM-Newton, an ESA science mission
with instruments and contributions directly funded by
ESA Member States and NASA.
R.S.X., H.L., \& Y.Q.X. acknowledge support from NSFC grants (12025303{, 12393814,} and 11890693), the Strategic Priority Research Program of the Chinese Academy of Sciences (grant NO. XDB0550300), the National Key R\&D Program of China (2022YFF0503401), and the science research grants from the China Manned Space Project with NO. CMS-CSST-2021-A06.
\end{acknowledgments}

\facilities{\textit{XMM/EPIC}}

\software{\textsc{astropy} \citep{2013A&A...558A..33A}, \textsc{emcee} \citep{2013PASP..125..306F}, \textsc{matplotlib} \citep{2013A&A...558A..33A}, \textsc{numpy} \citep{harris2020array}, \textsc{SAS} \citep{2004ASPC..314..759G}, \textsc{scipy} \citep{2020SciPy-NMeth}}

\bibliography{sample631}{}
\bibliographystyle{aasjournal}

\end{document}